\documentclass[10pt]{article}
\usepackage{natbib}
\usepackage{authblk}
\usepackage[textfont = {footnotesize} , labelfont = bf]{caption}
\usepackage{subfig}
\usepackage{graphicx}
\usepackage{setspace}
\def \iPAC{\emph{iPAC}}
\def \GraphPAC{\emph{GraphPAC}}
\def \NMC{\emph{NMC}}
\usepackage{url}  
\usepackage{ifthen}  
\usepackage{multicol}   
\usepackage{amsmath}
\usepackage{amsfonts}
\usepackage{natbib}
\usepackage{subfig}
\usepackage{graphicx}
\usepackage{epstopdf}
\usepackage[margin=1in]{geometry}
\begin{document}

\title{A Graph Theoretic Approach to Utilizing Protein Structure to Identify Non-Random Somatic Mutations}
 
\title{A Graph Theoretic Approach to Utilizing Protein Structure to Identify Non-Random Somatic Mutations}
\author[1]{Gregory Ryslik}
\author[2] {Yuwei Cheng}
\author[2,3]{Kei-Hoi Cheung}
\author[4]{Yorgo Modis}
\author[1,2]{Hongyu Zhao}

\affil[1]{Department of Biostatistics, Yale School of Public Health, New Haven, CT, USA}
\affil[2]{Program of Computational Biology and Bioinformatics, Yale University, New Haven, CT, USA}
\affil[3]{Yale Center for Medical Informatics, Yale School of Medicine, New Haven, CT, USA}
\affil[4]{Department of Molecular Biophysics \& Biochemistry, Yale University, New Haven, CT,USA}

\maketitle

\begin{abstract}
\noindent \textbf{Background:}
It is well known that the development of cancer is caused by the accumulation of somatic mutations within the genome. For oncogenes specifically, current research suggests that there is a small set of ``driver" mutations that are primarily responsible for tumorigenesis. Further, due to some recent pharmacological successes in treating these driver mutations and their resulting tumors, a variety of methods have been developed to identify potential driver mutations using methods such as machine learning and mutational clustering.  We propose a novel methodology that increases our power to identify mutational clusters by taking into account protein tertiary structure via a graph theoretical approach.

\noindent \textbf{Results:} We have designed and implemented \GraphPAC{} (\textbf{G}raph \textbf{P}rotein \textbf{A}mino acid \textbf{C}lustering) to identify mutational clustering while considering protein spatial structure. Using \GraphPAC{}, we are able to detect novel clusters in proteins that are known to exhibit mutation clustering as well as identify clusters in proteins without evidence of prior clustering based on current methods. Specifically, by utilizing the spatial information available in the Protein Data Bank (PDB) along with the mutational data in the Catalogue of Somatic Mutations in Cancer (COSMIC), \GraphPAC{} identifies new mutational clusters in well known  oncogenes such as EGFR and KRAS. Further, by utilizing graph theory to account for the tertiary structure, \GraphPAC{} identifies clusters in DPP4, NRP1 and other proteins not identified by existing methods. The R package is available at: \url{http://bioconductor.org/packages/release/bioc/html/GraphPAC.html}

\noindent \textbf{Conclusion:} \GraphPAC{} provides an alternative to \iPAC{} and an extension to current methodology when identifying potential activating driver mutations by utilizing a graph theoretic approach when considering protein tertiary structure.

\end{abstract}

\section{Background}
Cancer, one of the most widespread and heterogeneous diseases, is at its most fundamental level a disease brought on by the accumulation of somatic mutations \citep{vogelstein_cancer_2004}. These mutations typically occur in either tumor suppressors or oncogenes. While oncogenic mutations either tend to deregulate or up-regulate the resulting protein behavior, mutations within tumor suppressors typically lower the activity of genes that prevent cancer. Pharmacological intervention has shown to be more effective with inhibiting activating oncogenes than with restoring functionality of tumor suppressing genes. Combined with the theory of ``oncogene addiction", that many cancers are dependent upon a small set of key genes to drive their rapid cellular multiplication with the rest of the mutations simply being passenger mutations \citep{greenman_patterns_2007, weinstein_mechanisms_2006}, the identification of driver oncogenic mutations has become of critical importance in cancer research.


Due to the importance of this problem, several approaches have been proposed to detect naturally selected regions in which activating mutations occur. One general approach postulates that driver mutations will have a higher non-synonymous mutation rate as compared to the background level after normalizing for the length of the gene \citep{wang_prevalence_2002, bardelli_mutational_2003, sjoblom_consensus_2006}. Similarly, assuming that the neutral rate of nucleotide substitution is surpassed when positive selection is acting on a specific region, one can check if the ratio of nonsynonymous ($K_a$) to synonymous ($K_s$) mutations per site is greater than 1 \citep{kreitman_methods_2000}. Relatedly, \citet{ye_2010} and \citet{ryslik_2013} showed that mutational clusters can be indicative of activating mutations and that finding such clusters is a way to reduce the driver mutation search space needing to be analyzed. An alternative approach relies on creating classifiers to categorize mutations. Machine learning algorithms such as \emph{Polyphen-2} \citep{adzhubei_method_2010}, which predicts whether a missense mutation is damaging, and \emph{CHASM} \citep{carter_cancer-specific_2009}, which discriminates between known driver mutations and a set of passenger mutations, rely upon a set of rules developed using a variety of machine learning techniques such as Random Forests \citep{breiman_randomforest} and Support Vector Machines \citep{cortes_support-vector_1995}. These rules can be used to calculate a score for each mutation based upon both sequence and non-sequence-based features  such as evolutionary conservation, size and polarity of the substituted residue as well as accessible surface area \citep{reva_predicting_2011}. Other classifiers, such as \emph{SIFT} \citep{ng_predicting_2001}, use only a subset of these features, e.g. evolutionary conservation, for predictions.

While the methods based upon background mutational rates have had some success in identifying regions of positive selections or driver mutations, they nonetheless suffer from several shortcomings. First, many of these methods rely upon calculating the difference between synonymous and non-synonymous mutations but do not take into account that selection can act upon minute regions of the gene. Thus, when the mutations rates are averaged over the entire gene, the signal may be lost. Second, the methods proposed by \citet{kreitman_methods_2000}  and \citet{wang_prevalence_2002} do not differentiate between activating gain-of-function mutations and inactivating loss-of-function non-synonymous mutations. Third, many of the machine learning methods require an extensive rule set that must first be trained using a well annotated database that is still limited. Until the requisite literature and information is developed, the machine learning algorithm is unable to create a well-performing classifier. Furthermore, the rules must be updated periodically to reflect updated knowledge and information.

Building on the work of \citet{bardelli_mutational_2003} and \citet{torkamani_prediction_2008}, which stipulated that only a small number of specific mutations can activate a protein, \citet{ye_2010} developed Non-Random Mutational Clustering (\NMC{}) to identify potential activating mutations. \NMC{} works on the hypothesis that absent any previously known mutational hotspot, a mutational cluster is indicative of a possible activating mutation. For the null hypothesis that mutation locations are randomly located in a candidate protein represented in linear form, \NMC{} identifies clustering by evaluating whether there is statistical evidence of mutations occurring closer together on the line than expected by chance. While \NMC{} was able to implicate some cancer related genes, it is limited by the fact that it considers the protein as a linear sequence and does not take into account the tertiary protein structure. To account for protein structure information, \citet{ryslik_2013} developed \iPAC{} (\textbf{i}dentification of \textbf{P}rotein \textbf{A}mino acid \textbf{C}lustering), which reorganizes the protein into a one dimensional space that preserves, as best as possible, the three dimensional amino acid pairwise distances using Multidimensional Scaling (MDS)  \citep{borg_modern_1997}.  While it was shown that \iPAC{} provides an improvement over NMC, the reliance upon a global method like MDS can potentially result in a distorted rearrangement of the protein, since distant residues will nevertheless have an impact on each other's final position in one dimensional space.

In this manuscript, we provide an alternative method to \iPAC{} by remapping the protein into one dimensional space via a graph theoretic approach. This approach allows for a more natural consideration of the protein, one that is sensitive to protein domains and linkers. We show that our methodology is effective in identifying proteins with mutational clustering that are missed by both \iPAC{} and \NMC{} such as NRP1 and MAPK24. We also show that for some proteins, \GraphPAC{} identifies fewer clusters than inferred by both \iPAC{} and \NMC{} while for other proteins \GraphPAC{} identifies more clusters than the other two methods. While both \GraphPAC{} and \iPAC{} are an improvement over \NMC{} since they account for tertiary structure, the differences between \GraphPAC{} and \iPAC{} point to the fact that different rearrangements of the protein must be considered in order to better understand the mutational clustering landscape. We show that many of the clusters identified by \GraphPAC{} are also classified as damaging by \emph{Polyphen-2} and as an activating mutation by \emph{CHASM}. By providing a more complete picture of mutational clustering than \iPAC{} or \NMC individually, \GraphPAC{} allows us to obtain a more accurate landscape of where potential activating mutations may occur on the protein.

\section{Methods}

\GraphPAC{} uses a four step approach to identifying mutational clusters. The first step, as described in Sections \ref{data:COSMIC} and \ref{data:PDB}, retrieves mutational and positional data from COSMIC \citep{forbes_catalogue_2008} and the PDB \citep{pdb}, respectively. After reconciling the mutational and positional databases (Section \ref{data:Reconciliation}), the residues are realized as a connected graph where each residue is a vertex whereupon the traveling salesman problem is heuristically solved in order to find the shortest path through the protein (Section \ref{Calc:TSA}). Once the shortest path has been identified, the protein residues are reordered along this path providing a one dimensional ordering of the protein. The linear \NMC{} algorithm is then used to calculate which mutations are closer together than expected by chance. Lastly, the clusters are unmapped back into the original space and  the results reported back to the user. We detail each of the steps in the sections below.

\subsection{Obtaining Mutational Data} \label{data:COSMIC}

The mutational positions were obtained from the 58th version of the COSMIC database that was downloaded via the following ftp site: \url{ftp.sanger.ac.uk/pub/CGP/cosmic}. The database was implemented locally using Oracle 11g. Only missense mutations that were classified as ``Confirmed somatic variant" or ``Reported in another cancer sample as somatic" were selected, with nonsense and synonymous mutations excluded. Moreover, we only considered mutations originating from studies that were classified as whole gene screens. Next, since multiple studies can report mutational data from the same cell line, mutational redundancies were removed to avoid double counting the mutations. Lastly, only the proteins with a UniProt Accession Number \citep{the_uniprot_consortium_reorganizing_2011} were kept in order to correctly match the mutational and positional data, resulting in 777 proteins. See ``Cosmic Query" in the supplementary information for the SQL code required to generate the mutational data.

\subsection{Obtaining the 3D Structural Data} \label{data:PDB}
The PDB web interface was used to obtain the protein tertiary information for each of the 777 proteins described in Section \ref{data:COSMIC}. Since multiple structures are often available for the same protein, all structures with a matching UniProt Accession Number were used and an appropriate multiple comparisons adjustment (see Section \ref{MultCompStruct}) was performed afterwards.  For proteins where the resolution provided alternative conformations, the first conformation listed in the file was used. Similarly, for structures where more than one polypeptide chain with a matching Uniprot Accession Number was available, the first matching chain listed in the file was used (typically chain A). Finally, after the side-chain and conformation are selected, the cartesian coordinates of all the $\alpha$-carbon atoms are used to represent the tertiary backbone structure of the protein. See ``Structure Files" in the supplementary materials for a full listing of all the 1,904 structure/side chain combinations used.

\subsection{Reconciling the Structural and Mutational Data} \label{data:Reconciliation}
In order to reference the same residue in the COSMIC and PDB databases, an alignment was performed to accommodate their different numbering systems. Like \iPAC{}, \GraphPAC{} allows two such reconciliations. The first is based upon a pairwise alignment as described in \citet{Biostrings_2012} while the second is based upon a numerical reconstruction from the structural information available in the PDB file. Due to the fact that the PDB file structure potentially changes depending upon the structure release date along with other technical complications, pairwise alignment was used for all the analysis described in this paper unless specifically noted. For further information on the alignment please see the documentation in the \GraphPAC{} package available on Bioconductor. Protein/structure/side-chain combinations that resulted in only one mutation or no mutations on the residues for which tertiary information was available were dropped. Similar to \iPAC{}, a successful alignment of the tertiary and mutational data was obtained for 140 proteins corresponding to 1100 unique structure/side-chain combinations. See ``Structure Files" in the supplementary materials for a full listing and description.

\subsection{Traveling Salesman Approach} \label{Calc:TSA}
\begin{figure}[h!]
	\centering
	\includegraphics[scale=0.5]{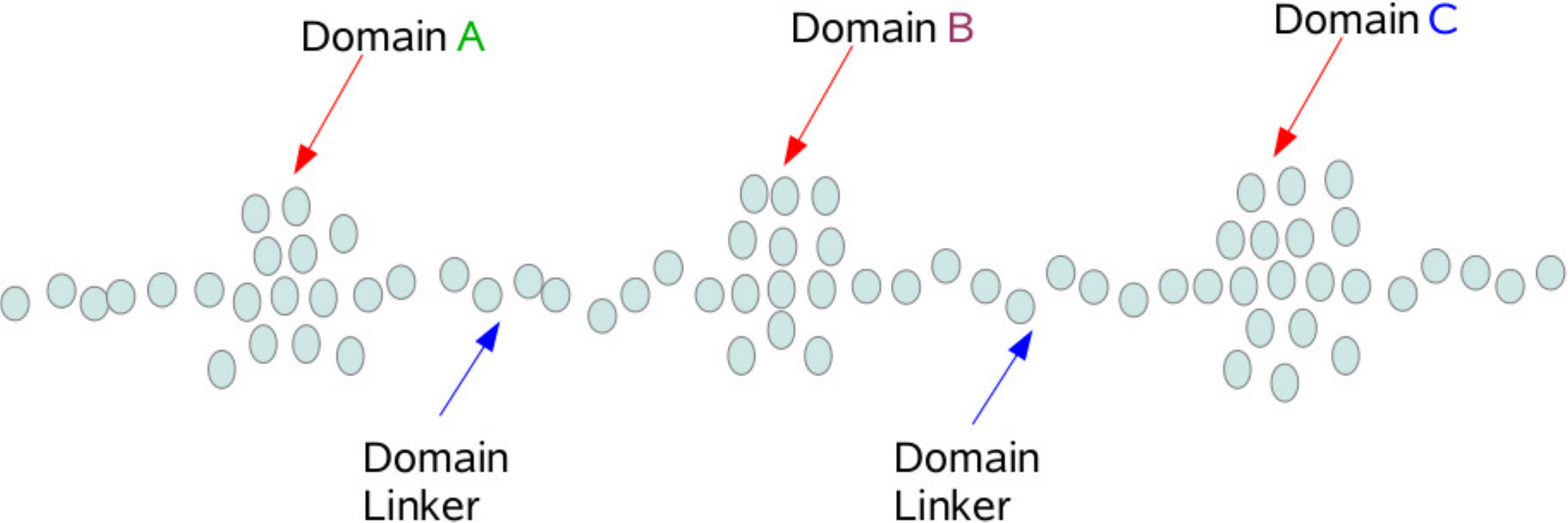}
	\caption{An example protein with three different domains. Under \iPAC{}, the residues in Domain A will have an effect on the final position on the residues in Domain C and vice versa, a result that is undesirable if the three domains are independent of each other. The residues in Domain A and Domain C will have no effect on each other's final position via the graph theoretic approach.}
	\label{fig:Domains}
\end{figure}

Since the \NMC{} algorithm requires order statistics to identify clustering (see Section \ref{Calc:NMC}), we need to map the protein from a three dimensional to a one dimensional space so that order statistics may be constructed. Contrary to \iPAC, which employed MDS, a graph theoretic approach is used by \GraphPAC{}. As discussed above, one major limitation of MDS is that the minimization of the stress function: \begin{eqnarray} \label{stress}
\sigma_1 = \sqrt{\frac{\sum_{i,j}[f(\delta_{i,j}) - d_{i,j}(\mathbf{X})]^2}{\sum_{i,j}d_{i,j}^2(\mathbf{X})}}
\end{eqnarray} results in every residue having an effect on the final position of every other residue.  In Equation \ref{stress}, $\delta_{ij}$ represents the Euclidean distance between residues $i$ and $j$ in the original higher-dimensional space while $d_{i,j}(\mathbf{X})$ represents the distance between them in the lower dimensional space $\mathbf{X}$. Lastly, $f:\delta_{i,j} \rightarrow d_{i,j}(\mathbf{X})$, is used to account for situations where the proximity measures $\delta_{i,j}$ do not come from a true metric space. Since in our case, $\delta_{i,j} \in \mathbb{R}^3$, $f$ is the identity function. Minimization of $\sigma_1$ may not capture that a protein is typically comprised of several domains and that only residues within a specific domain should influence each other's final position in linear space (see Figure \ref{fig:Domains}).

Under the \GraphPAC{} algorithm, we first construct a complete graph with each residue represented by a vertex. We then create a linear ordering of the protein by finding a Hamiltonian\footnote{A Hamiltonian path is a walk through the graph that visits every vertex once and only once.} path through the graph. As the number of distinct Hamiltonian paths on a graph with $N$ vertices is equal to $\frac{(N-1)!}{2}$, a direct consideration of all possible paths is computationally unfeasible. Further, selective pruning of the edges based upon edge distance is also often impractical due to the domain structure where  many residues are close to each other. Because of these factors, we use a heuristic algorithm that solves the Traveling Salesman Problem (TSP) \citep{applegate_traveling_2006, TSP_package_2007} to find a linear path that is approximate of the shortest path through the protein. We then use this path as a representative reordering of the protein into one dimensional space to identify clusters. Unlike \iPAC{}, whic is based on a global remapping, this methodology takes into account only locally neighboring residues to remap the protein to one dimensional space. 

While there are many heuristic solutions for the TSP (see \citet{gutin_traveling_2007}), we consider three of the most common insertion methods \citep{rosenkrantz_analysis_1977}: cheapest insertion, farthest insertion and nearest insertion as described below. Specifically, the objective of the TSP is to find a  cyclic permutation $\pi$ of $\{1,2,3, \dots , n\}$ that minimizes the total tour distance, namely:
$$\min_\pi \sum_{i=1}^n d(i, \pi(i))$$ Here, $d(i,j)$ represents the distance between residues $i$ and $j$ (with $d(i,i) = 0$) and $\pi(i)$ represents the residue that follows residue $i$ on the tour.  The difference between the three insertion methods rests on how the next residue $k$ is selected for insertion. Under cheapest insertion, the next $k$ to be inserted into the tour is chosen such that the increase in tour length is minimal. Under nearest insertion, at each iteration, the $k$ that is closest to a residue already on the tour is selected. Finally, under farthest insertion, the $k$ that is farthest away from any residue already on the tour is selected.

These algorithms have different upper bounds on their tour lengths. For example, the farthest insertion algorithm creates tours that approach $\frac{3}{2}$ of the shortest length while the nearest and cheapest insertion algorithms can be linked to the minimal spanning tree algorithm and thus have an upper bound of twice the shortest tour length when distances satisfy the triangular inequality \citep{TSP_package_2007}. Due to the varied nature of these methods and that there is no biological justification to favor one over the other, we consider all three methods when identifying clusters and then perform an appropriate multiple comparison adjustment to infer the statistical evidence of mutation clusters (see Section \ref{MultCompStruct}). 

\begin{figure}[h!]
	\centering
	\includegraphics[scale=0.30]{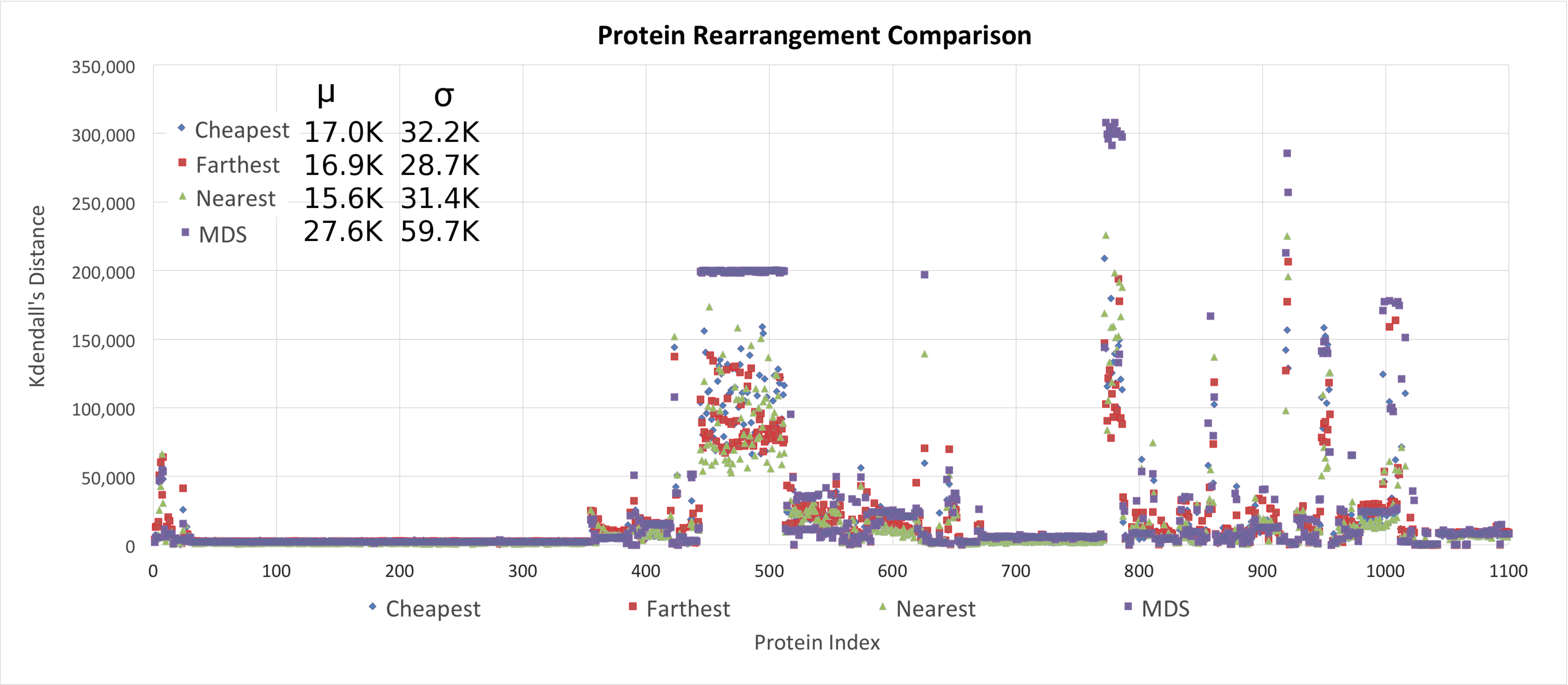}
	\caption{The amount of rearrangement performed under each of the three insertion methods described as well as MDS. Each column on the x-axis represents one of the 1100 structures considered, with structures from the same protein adjacent to one another and the protein order determined lexicographically by protein name. The y-axis shows the Kendall Tau distance, which is equivalent to the number of swaps required to sort the protein back into \{1,2,3,\dots,..\} order using bubble sort. The proteins with at least one rearrangement higher than 150,000 represent the DPP4, F5, IDE, MET, PIK3C$\alpha$, SEC23A and TF proteins, from left to right, respectively.}
		\label{fig:Distribution}
\end{figure}

As can be seen from Figure \ref{fig:Distribution}, all the rearrangement options present a positive skew and are mostly consistent with each other. For the majority of the proteins, all three insertion approaches as well as the MDS approach result in little rearrangement. However, if one method results in radical rearrangement when the protein is mapped to 1D space, the other methods do so as well. This makes selection of a specific insertion method less critical and for the rest of this manuscript, unless otherwise specified, we use the insertion method with the most significant cluster for analysis. Please see ``Distribution Summary" in the supplementary materials for a full listing of each structure's Kendall Tau distance, protein index and a high resolution plot.

\subsection{NMC} \label{Calc:NMC}

\begin{figure}[h!]
	\centering
	\includegraphics[scale=0.50]{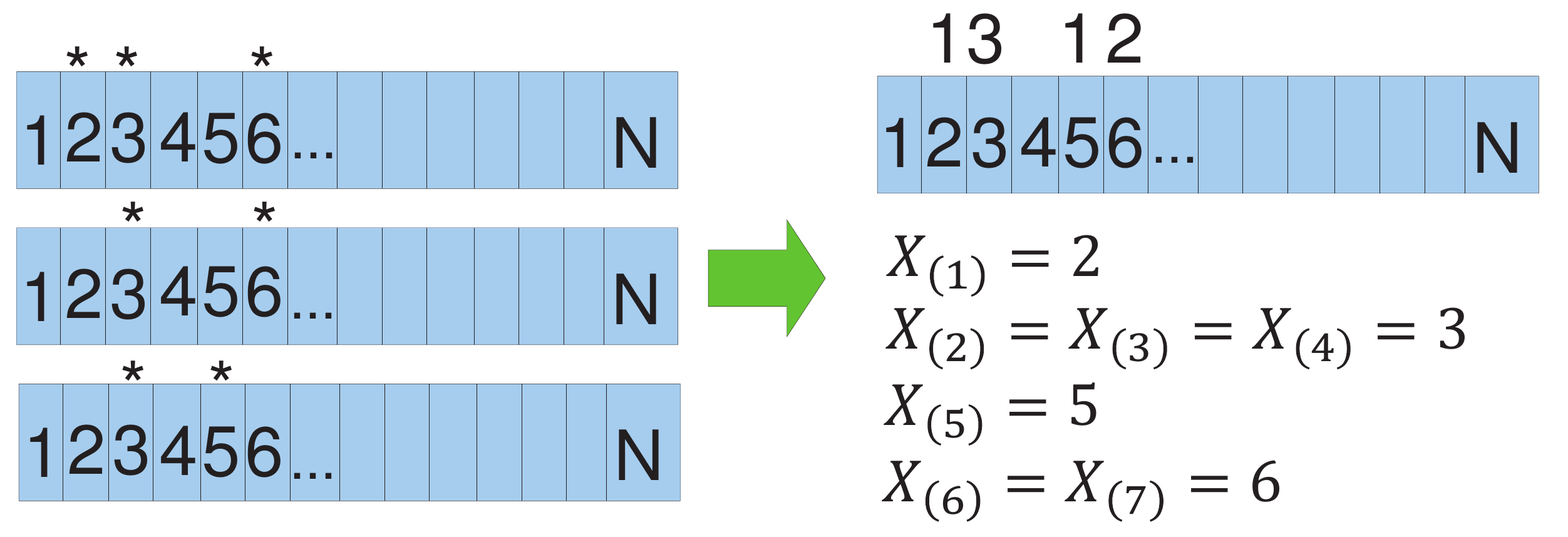}
	\caption{An example constructing order statistics over 3 samples with 7 total mutations. The number inside the box indicates the residue number. A ``*" above a residue signifies a non-synonymous missense substitution mutation for that residue. Figure from \citet{ryslik_2013}.}
	\label{fig:OrderStats}
\end{figure}

The \NMC{} algorithm as described by \citet{ye_2010}, and briefly reviewed here, was used to find the mutational clusters once the protein was remapped to 1D space. To begin, suppose we had $m$ samples of a protein that was $N$ residues long and that there were a total of $n$ mutations over all $m$ proteins. As shown in Figure \ref{fig:OrderStats}, by collapsing over the $m$ samples, we can construct order statistics for every mutations. Then, given order statistics $X_{(k)}$ and $X_{(i)}$ where $i < k$, we define a cluster to exist if $Pr(C_{ki} = X_{(k)} - X{(i)}) \leq \alpha$, for some predetermined significance level $\alpha$. As shown in \citet{ye_2010}, while a closed form calculation of the above probability is possible, it often becomes computationally costly. To overcome this, we calculate $\frac{C_{ki}}{N}$ and assume that the statistic is uniform on $(0,1)$.  Then in limit, it can be shown that:

\begin{equation} \label{eqn:continuous}
\begin{aligned}
&Pr(\frac{C_{ki}}{N}=\frac{X_{(k)} - X_{(i)}}{N} \leq c) \\
&= \int_0^c \frac{n!}{(k-i-1)!(i+n-k)!} y^{k-i-1}(1-y)^{i+n-k} dy\\
&= Pr(Beta(k-i, i+n-k+1) \leq c)
\end{aligned}
\end{equation}


The above calculation is then performed on all pairwise mutations and an appropriate multiple comparison adjustment is then applied. For the remainder of this study, we use the more conservative Bonferroni correction \citep{Bonferroni_1959, Bonferroni_1961} to adjust for the intra-protein cluster p-values. See Section \ref{MultCompStruct} for a description of how we account for the inter-protein multiple comparisons. Lastly, it is important to mention that the structural information obtained for each protein does not always contain the $(x,y,z)$ coordinates for every residue in the protein. In such cases, in order to compare \GraphPAC, \iPAC{} and \NMC{} on an equal basis, these missing residues are removed from the protein. 

\subsection{Multiple Comparison Adjustment For Structures} \label{MultCompStruct}
In addition to the Bonferroni adjustment performed to account for multiple testing within a specific structure, we perform a second multiple comparison adjustment to account for testing all 1100 structures. Since a single protein can have many structures that are similar to each other, a second Bonferroni adjustment is too conservative and an integrated Bonferroni-FDR approach was performed. Specifically, for a given protein, the Bonferrroni adjusted p-value of each cluster was multiplied by $\frac{n(n-1)}{2}$ to calculate $p^*$. Thus, $p^*$ could be compared directly to an $\alpha$-level of $0.05$ in order to determine the cluster's significance. Next, a rFDR\citep{gong_atlas_2009} approach, which is a good approximation for the standard FDR method when there are a large number of independent or positively correlated tests, was used. Under this method, the expected value of $\alpha$ is estimated over all $k$ tests and then used as the significance threshold. Setting $k$ as the total number of structures under all three insertion methods, the mean alpha can be approximated by: $$rFDR = \alpha\left(\frac{k+1}{2k}\right)$$ where $k=3\times 1100=3300$. Using $\alpha = 0.05$, $rFDR$ is calculated to be $\approx 0.025007$. Rounding down, all the clusters for which $p^* \leq 0.025$ were deemed to be significant. To avoid confusion in the rest of the paper, we only report the p-value (with the exception of Table \ref{tab:8Tie}). However, each cluster discussed in Section \ref{discussion} is significant after the Bonferroni-FDR multiple comparison adjustment described here.

\section{Results and discussion}\label{discussion}

Using the \GraphPAC{} algorithm, out of the 140 proteins analyzed, 9, 10 and 12 significant proteins were found under the cheapest, nearest and farthest insertion methods, respectively. This corresponded to 223, 225 and 226 significant structures (out of the 1100 total structures considered) under the three methods. Eight proteins were identified as having significant clusters by all three insertion methods of \GraphPAC{}, as well as \NMC{} and \iPAC{}, see Table \ref{tab:8Tie}. Compared to NMC, five additional proteins were identified when local spatial structures were considered: EGFR (nearest insertion), DPP4 (farthest insertion), MAP2K4 (cheapest and nearest insertions), NRP1 (farthest insertion) and PCSK9 (farthest insertion). Among these 5 proteins, iPAC, which uses global spatial structure, only identified EGFR as having mutation clusters. These 5 proteins correspond to a total of 6 structures, with two structures having significant clustering for EGFR. See Section \ref{relevant} for a summary of cluster overlap with active biological sites along with performance evaluation via machine learning methods.
 It is important to note, that there were no proteins found to have significant clustering under the linear \NMC{} algorithm that were subsequently missed by the \GraphPAC{} algorithm.
Further, \GraphPAC{} identified four proteins with clustering that are missed by the \iPAC{} algorithm: DPP4, MAP2K4, NRP1, and PSCK9. DPP4 is a serine protease that can modify tumor cell behavior and is a potential cancer therapeutic target \citep{kelly_fibroblast_2005}. Both MAP2K4 and NRP1 are well known to be associated with lung cancer \citep{ahn_map2k4_2011, lantuejoul_expression_2003}. Finally, while PCSK9 mutations are well known in causing hypercholesterolemia \citep{abifadel_mutations_2003}, recent research shows that absence of PCSK9 can provide a protective benefit against melanoma due to lower circulating LDLc in turn providing a potential cancer therapy via PCSK9 inhibitors \citep{seidah_proprotein_2013}. For a full listing of which structure-protein combinations were found significant, see ``Results Summary" in the supplementary materials.

\begin{figure}[h!]
	\centering
	\includegraphics[scale=0.42]{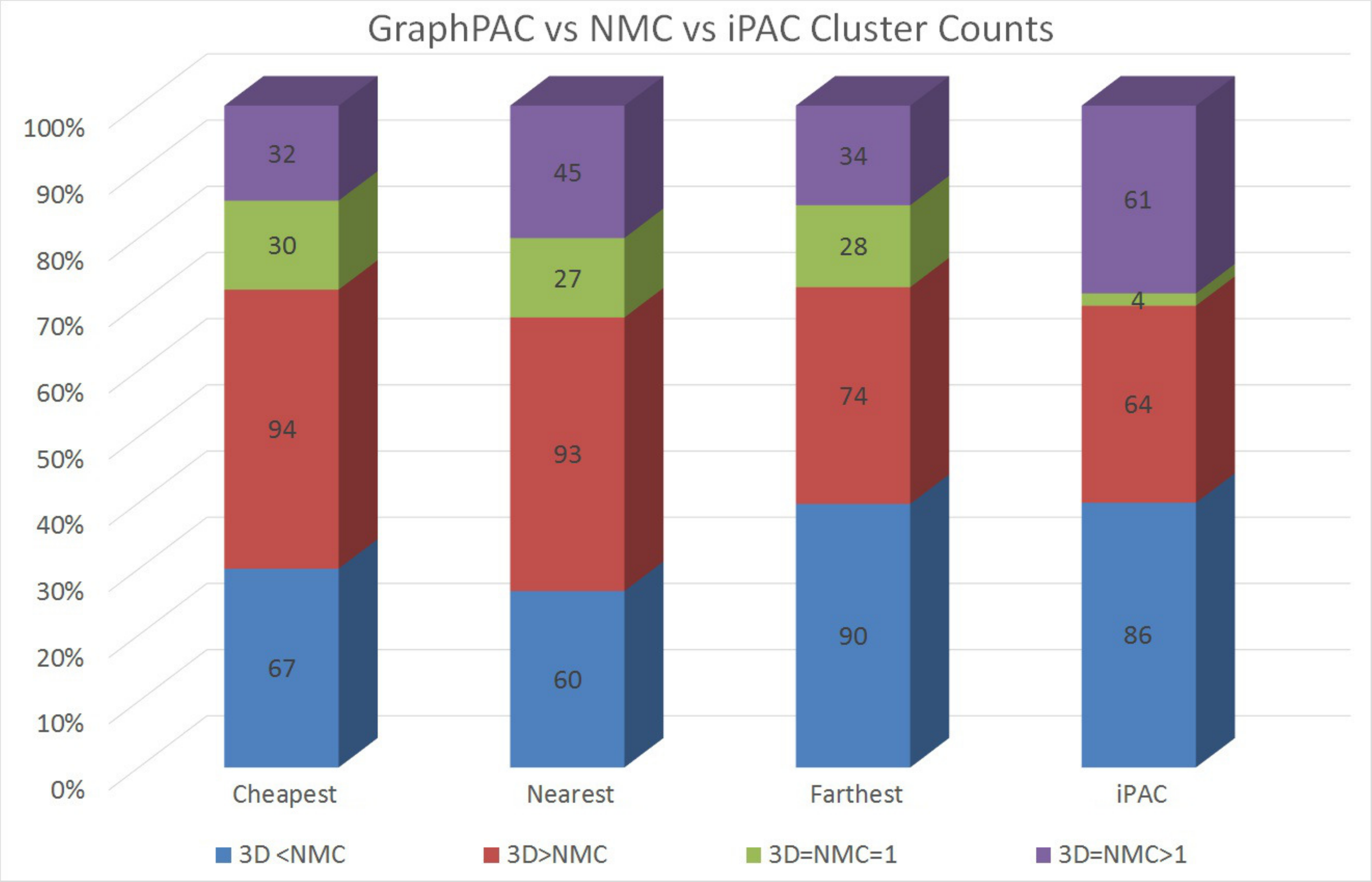}
	\caption{A comparison of \GraphPAC{}, \iPAC{} and \NMC{} over all the structures that were found to be significant. Each of the 3D methods are considered: all three \GraphPAC{} insertion methods and \iPAC{}. The size of each colored block represents the number of structures with the relationship described. For instance, the bottom blue block shows that 67 (of the total 223) significant structures using the \GraphPAC{} cheapest insertion method had fewer clusters as compared to the \emph{NMC} method.}
	\label{fig:StructureTally}
\end{figure}

As shown for each of the methods in Figure \ref{fig:StructureTally}, failure to utilize the tertiary information results in either an over or an underestimation of the number of clusters in approximately 70\% of the structures analyzed. Hence, failure to account for the protein structure provides either an overly complicated or overly simplified view of the mutational orientation. Please see Sections \ref{new}, \ref{more} and \ref{less} for an in-depth review of selected protein-structure combinations.

\begin{table}[H!tb]
\centering

    \begin{tabular}{|c|c|c|c|c|}
        \hline
        ~            &\multicolumn{2}{c|}{\GraphPAC{}} &  \multicolumn{2}{c|}{\emph{NMC}}   \\ \hline
 Protein &  p-value & p*  & p-value & p*\\ \hline
 KRAS   &  4.21 E-233 & 4.33 E-229 & 4.39 E-233 & 4.52 E-229\\       
 TP53   &  4.05 E-152 & 4.48 E-147 & 4.37 E-086 & 5.30 E-81\\    
 BRAF   &  3.84 E-130 & 1.04 E-126 & 3.84 E-130 & 1.04 E-126 \\       
 PIK3CA &  8.20 E-084 & 3.58 E-080  & 8.20 E-084 & 3.58 E-080\\ 
 NRAS   &  8.26 E-029 & 9.91 E-027  & 8.26 E-029 & 9.91 E-027\\ 
 HRAS   &  1.54 E-014 & 6.94 E-013  & 5.61 E-010 & 8.42 E-009 \\                        
 AKT1   &  2.47 E-005 & 2.47 E-004  & 2.47 E-005 & 7.41 E-005  \\ 
 IDE    &  1.56 E-003 & 4.67 E-003  & 1.56 E-003 & 4.67 E-003
\\ 
    \hline
    \end{tabular}

\normalsize
    \vspace{4 pt}
	\caption{A comparison of 8 proteins that were found to be significant by both \GraphPAC{} and NMC. The p* calculation is described in section \ref{MultCompStruct}. The smallest p-value from all of the insertion methods was selected.}
	\label{tab:8Tie}
\end{table}

\subsection{Cluster localization in relevant sites and perforamance evaluation} \label{relevant}
We note that 9 of the 13 proteins that \GraphPAC{} identified as having significant clustering have their most significant cluster overlap a binding site, catalytic domain or kinase domain. Out of the remaining four proteins, three proteins have their most significant cluster fall within a previously identified biologically relevant region. For instance, IDE's most significant cluster is located on residues 684-698, a denaturation-resistant epitope region \citep{cavender_transactivation_1999}. For NRP1, which plays roles in angiogenesis \citep{jubb_neuropilin-1_2012} and axon guidance \citep{maden_nrp1_2012}, the most significant cluster directly overlaps the F5/8 type C 1 domain - a domain in many blood coagulation factors. Finally, for PIK3C-$\alpha$, the most significant cluster overlaps residue 1047 which has been shown to potentially increase the substrate turnover rate, a common oncogenic behavior \citep{mankoo_pik3ca_2009}. For further detail on relevant biological site information, please see ``Relevant Sites" in the supplementary materials.

Further, we evaluated the performance of \GraphPAC{} via two well-known machine learning algorithms: \emph{CHASM} \citep{carter_cancer-specific_2009} and \emph{PolyPhen-2} \citep{adzhubei_method_2010}. It is critical to first note however, that the machine learning algorithms utilize a much more detailed set of features when evaluating the mutation. Thus these algorithms may identify mutations as significant while \GraphPAC{} would not. Nevertheless, of all the mutations that fall within significant clusters identified by \GraphPAC{}, 93\% and 91\% of them were also identified as significant (FDR $\leq$ 20\%) by \emph{CHASM} and $\emph{PolyPhen-2}$ (respectively). The benefit of \GraphPAC{} is that it is able to be executed with far less prior information. For further details, see ``Performance Evaluation" in the supplementary materials.

\subsection{\GraphPAC{} finds novel proteins compared to \iPAC{} and \emph{NMC}} \label{new}
As described in Section \ref{discussion}, \GraphPAC{} identified five additional proteins as compared to the linear \NMC{} algorithm. In this section we will consider two of these proteins which are both directly cancer related: EGFR, which is also identified by \iPAC{}, and NRP1, which is not identified by \iPAC{}.

\begin{figure}[h!]
	\centering
	\includegraphics[scale=0.15]{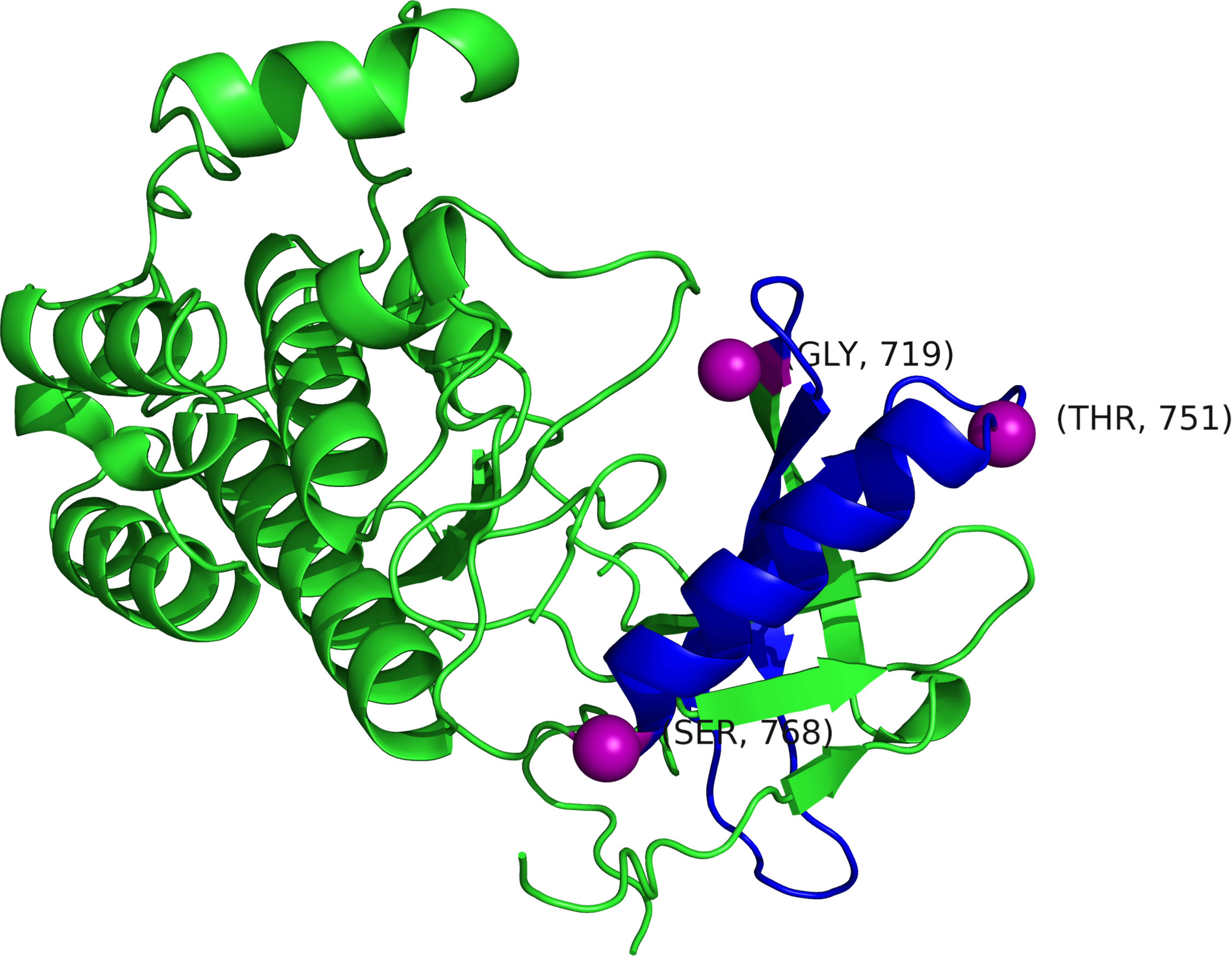}
	\caption{The EGFR ectodomain fragment structure (PDB ID 2ITX) where the 719-768 cluster is colored in blue. The three mutations, 719,751 and 768 are displayed as purple spheres. }
	\label{fig:EGFR}
\end{figure}

EGFR is a cell-surface receptor for ligands in the epidermal growth factor family \citep{Herbst2004S21} and is present in a wide range of diseases such as glioblastoma multiforme \citep{Heimberger15022005}, lung adenocarcinoma \citep{ladanyi_lung_2008} and colorectal cancer \citep{markman_egfr_2010}. The most significant cluster found was in the 2ITX structure \citep{pdb_2ITX} between residues 719-768 (see Figure \ref{fig:EGFR}) with a corresponding p-value of 0.0009. This cluster contains mutations G719S, T751I and S768I which are all found in non-small cell lung carcinomas (NSCLC) \citep{yoshikawa_structural_2012,simonetti_detection_2010,masago_good_2010} with mutation G719S well known for increased kinase activity \citep{yun_structures_2007}. It is also interesting to note that all three mutations within this cluster, which was identified purely through statistical clustering analysis, show a beneficial clinical response to either Erlotonib or Getfinib \citep{kancha_epidermal_2011,peraldo-neia_epidermal_2011,masago_good_2010}. Exclusion of the tertiary information would have resulted in this cluster being missed.

\begin{figure}[h!]
	\centering
	\includegraphics[scale=0.15]{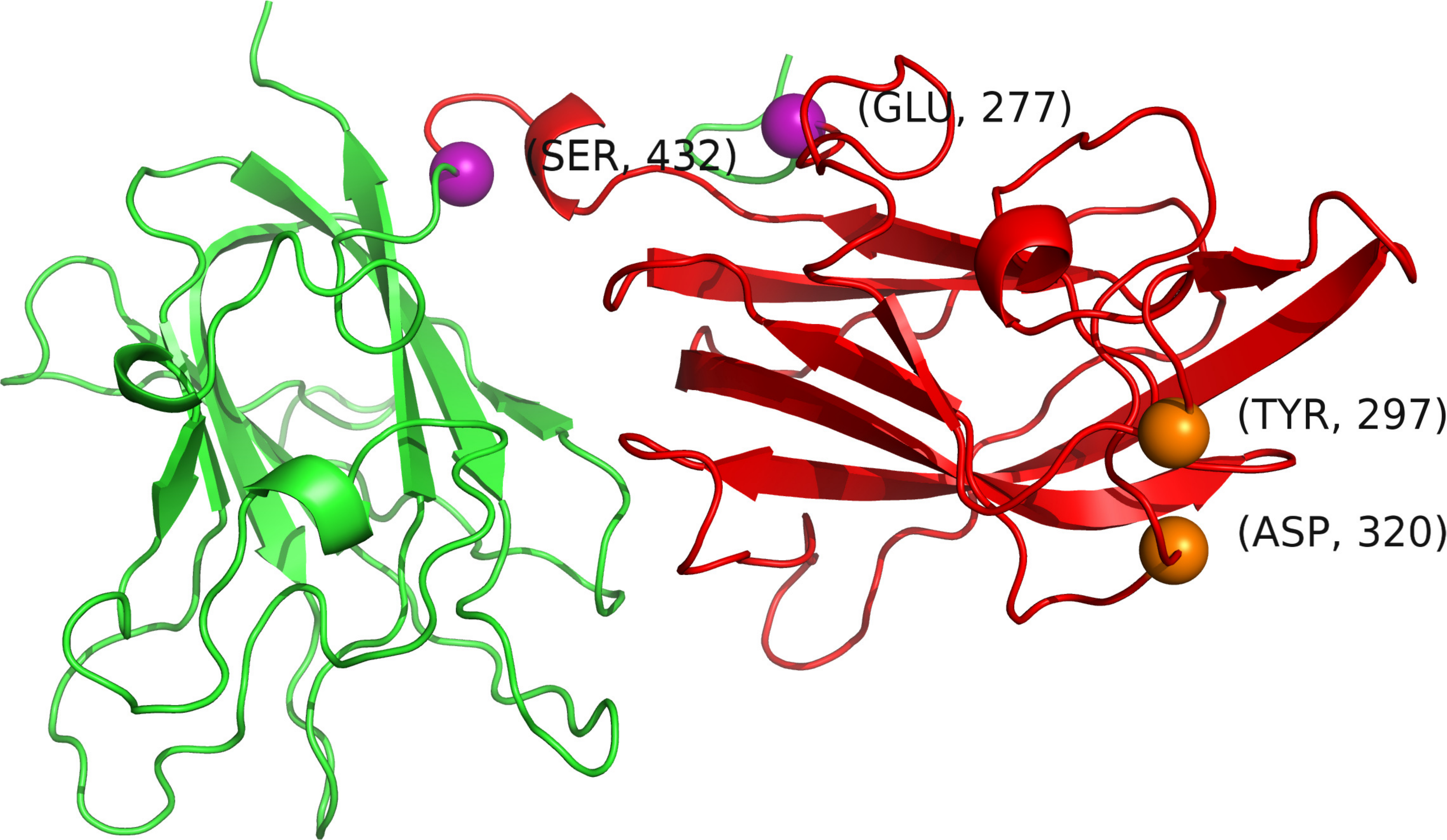}
	\caption{The NRP-1 structure (PDB ID 2QQI) structure where the 277-432 cluster is colored in red. The mutations that disrupt VEGF binding, 297 and 320 are shown as orange spheres while the end-points of the cluster, 277 and 432, are shown in purple spheres. }
	\label{fig:NRP1}
\end{figure}

We now consider the NRP-1 protein, a coreceptor for the vascular endothelial growth factor (VEGF) which is upregulated in a large variety of cancers including lung tumors \citep{lantuejoul_expression_2003}, gastrointestinal metasteses \citep{hansel_expression_2004} and pancreatic carcinomas \citep{parikh_expression_2003}. In NSCLC patients, it has been shown to be an independent predictor of cancer relapse and reduced survival as well as a cancer invasion enhancer \citep{hong_targeting_2007}. Moreover, research has shown that NRP-1 inhibitors provide an additive effect to anti-VEGF therapy in reducing tumor progression. Monoclonal antibodies that attach to the b1-b2 domains, the domains responsible for VEGF binding, have already been created \citep{pan_blocking_2007}. The b1 domain, which spans residues 275-424 almost exactly overlaps the most significant cluster found by \GraphPAC{}, which consists of residues 277-432 (p-value 0.0158) in the 2QQI \citep{pdb_2QQI} structure (Figure \ref{fig:NRP1}). 
Finally, it is worth noting that mutations on residues 297 and 320 were recently found that completely disrupt VEGF binding, both of which also fall within the \GraphPAC{} identified cluster of 277-432 in the 2QQI structure. 

\subsection{\GraphPAC{} identifies additional clusters compared to \iPAC{} and \emph{NMC}} \label{more}
\begin{figure}[h!]
	\centering
	\includegraphics[scale=0.15]{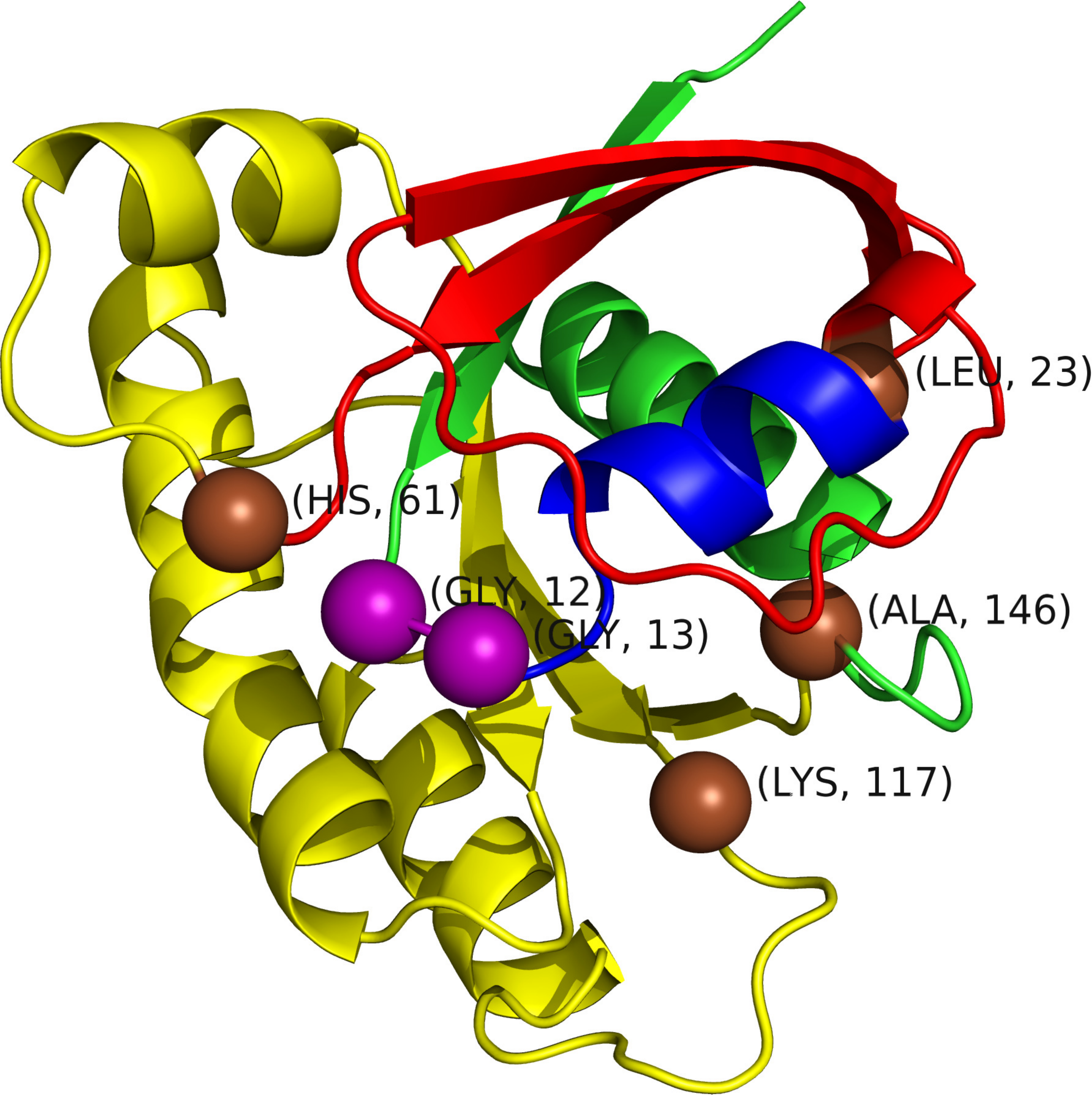}
	\caption{The KRAS structure (PDB ID 3GFT) color coded by region: amino acids 13-22 are blue, 24-60 are red and 62-145 are yellow. Residues 12 and 13 which make up the most significant cluster are shown as purple spheres, while residues 23, 61, 117 and 146 are shown as brown spheres. }
	\label{fig:KRAS-3GFT}
\end{figure}

A representative example where \GraphPAC{} identifies additional clusters as compared to \NMC{} and \iPAC{} is in the KRAS protein for the 3GFT structure \footnote{For this analysis, a manual reconstruction was performed in order to include residue 61 which is listed as a histidine under isoform 2B in the Uniprot Database and a glutamine in the COSMIC database. As the substitution of one amino acid in the structure would not have a significant impact on the spatial structure of the protein, and residue 61 is a highly mutated position, the residue was kept in the analysis. As a result, amino acids 1 - 167 are used.}\citep{kras_3GFT} (Figure \ref{fig:KRAS-3GFT}). KRAS, a GTPase,  is one of the most pervasively activated oncogenes, with some estimates stating that between 17-25\% of all human tumors contain an activating mutation of the gene \citep{Kranenburg_2005}. Due to the large number of samples with mutations in this gene and the resulting strong statistical signal, \GraphPAC{}, \iPAC{} and \NMC{} all identify that KRAS contains highly statistically significant mutational clusters. Nevertheless, \GraphPAC{} identifies several novel clusters that are missed by \iPAC{} and NMC. While all three methods identify clustering at residues 12-13, 12-61 and 12-146, only \iPAC{} and \GraphPAC{} identify two additional clusters at 1) 61-117 and 2) 117-146.

\begin{table} [H!tbp]
	\begin{center}
    \begin{tabular}{|c|c|c|c|}
        \hline
        ~        & ~          & ~          & ~           \\ \hline
        Residues & \NMC{}        & \iPAC{}    & \GraphPAC{} \\ 
        12-13    & 9.45 E-229 & 3.91 E-165 & 8.95 E-229  \\ 
        12-23    & -          & -          & 1.31 E-99   \\ 
        12-61    & 4.34 E-65  & 2.38E E-87 & 5.49 E-164  \\ 
        12-146   & 3.85 E-13  & 3.81 E-90  & 2.87 E-16   \\ 
        23-61    & -          & -          & 1.01 E-105 \\ 
        61-146   & -          & 3.01 E-106 & 4.35 E-31   \\ 
        117-146  & -          & 1.66 E-102 & -           \\ 
        \hline
    \end{tabular}
	\end{center}
	 \caption{P-value comparison of the three algorithms for several significant clusters. A ``-" signifies that the method did not find that cluster to be significant. For \GraphPAC{}, the cheapest insertion results are reported here.}
	 	\label{tab:3GFT}
\end{table}

Moreover, only \GraphPAC{} (under the cheapest and nearest insertion methods) identifies a statistically significant cluster for residues 12-23 and 23-61 as shown in Table \ref{tab:3GFT}. Considering the 12-23 cluster, we see that a sub-cluster of 12-13 is identified as well. This follows biological function as mutations on residues 12 and 13 appear in a large variety of cancers, such as breast, lung, bladder, pancreas and colon \citep{mccoy_human_1984, motojima_detection_1993, sjoblom_consensus_2006}  while mutations on residues 22 and 23 appeared in colorectal/large intestine tissue samples in our data. It is interesting to note that germline mutations on residue 22  often result in developmental disorders such as Noonan Syndrome Type 3 (NS3) as well as Cardiofaciocutaneous Syndrome (CFC) \citep{zenker_expansion_2007, gremer_germline_2011}. 

Finally, the majority of mutations in cluster 61-146 also fall along biological lines with all the mutations in our data either occurring in lung or gastrointestinal tract/large intestine carcinomas. Specifically, residue 61 is highly mutable with mutations found in colorectal and lung cancer \citep{sjoblom_consensus_2006, tam_distinct_2006} while mutations K117N and A146T are found specifically in colorectal cancer \citep{sjoblom_consensus_2006}.

\subsection{\GraphPAC{} finds fewer clusters compared to NMC} \label{less}
\begin{figure}[h!]
	\centering
	\includegraphics[scale=0.15]{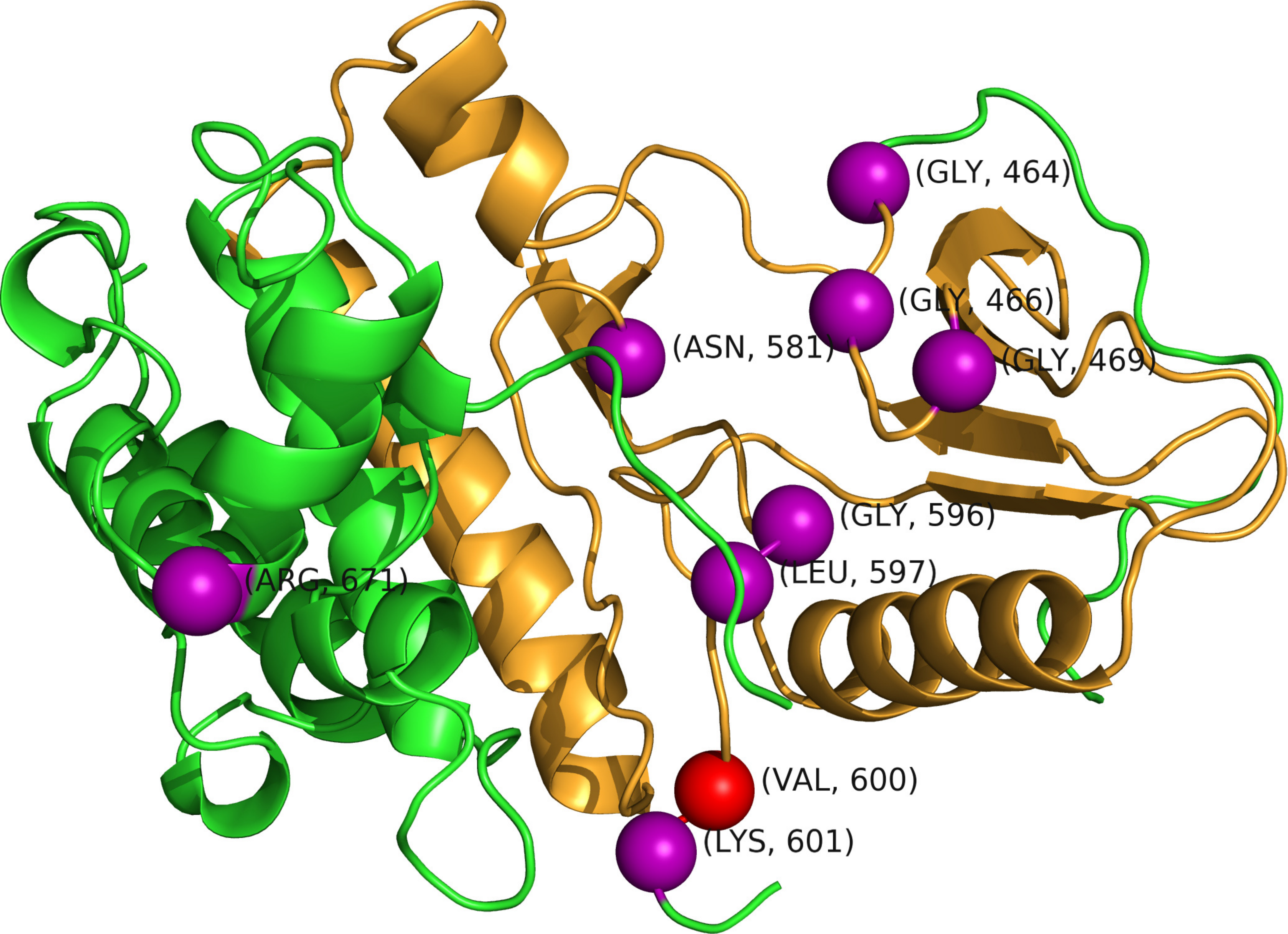}
	\caption{The BRAF structure (PDB ID 4E26) color coded by segment: I) amino 464-599 are orange 2) amino acids 601-671 are green. The $\alpha$-carbons of the mutated residues 464, 466, 469, 581 ,596, 597 , 601 and 671 are shown as purple spheres. Residue 600 is shown as a red sphere.}
	\label{fig:BRAF-3TV4}
\end{figure}

As seen from Figure \ref{fig:StructureTally}, between 25\%-40\% of the structures identified with significant clustering had fewer clusters under the \GraphPAC{} methodology as compared to the linear \NMC{} algorithm with the vast majority of these structures corresponding to BRAF, HRAS and TP53. Here we consider a representative example, the 4E26 structure \citep{pdb_4E26} for BRAF when analyzed using the farthest insertion method (Figure \ref{fig:BRAF-3TV4}). As \iPAC{} identified even more clusters than NMC, we compare \GraphPAC{} to NMC in this section when showing that fewer mutational clusters is of benefit. Further, as V600 is well known to be the most likely mutated position in BRAF, the most significant cluster identified by \GraphPAC{}, \iPAC{} and \NMC{} is located only on that residue with a p-value of $2.12 \times 10^{-129}$ under all three methods. In all, \GraphPAC{} identifies 16 clusters while \NMC{} identifies 22, with the differences shown in Table \ref{tab:BRAF}.

\begin{table}[h!tbp]
\centering
\subfloat[Clusters found by  \GraphPAC{}. A "-" for the \NMC{} value signifies that cluster was not identified under the linear algorithm.]{
    \begin{tabular}{|c|c|c|c|c|}
        \hline
        ~     & ~   & ~                & \multicolumn{2}{c|}{p-value}                      \\ 
        Start & End & \# Muts. & \GraphPAC{}               & \NMC{}                     \\ \hline       
	600 & 600 & 60 & 2.12E-129 & 2.12E-129\\ 
	597 & 600 & 62 & 1.49E-104 & 1.49E-104\\ 
	600 & 601 & 62 & 1.49E-104 & 9.22E-117\\ 
	596 & 600 & 64 & 7.16E-102 & 7.16E-102\\ 
	596 & 601 & 66 & 3.37E-91 & 1.16E-100\\ 
	597 & 601 & 64 & 8.07E-91 & 7.16E-102\\ 
	601 & 671 & 3 & 5.85E-38 & -\\ 
	600 & 671 & 63 & 8.30E-37 & 7.08E-26\\ 
	469 & 601 & 72 & 2.59E-22 & 5.92E-17\\ 
	581 & 601 & 68 & 1.23E-21 & 1.33E-65\\ 
	581 & 600 & 66 & 2.94E-20 & 3.13E-63\\ 
	469 & 600 & 70 & 3.98E-20 & 4.91E-15\\ 
	466 & 601 & 74 & 2.15E-17 & 9.69E-19\\ 
	466 & 600 & 72 & 7.01E-16 & 1.60E-16\\ 
	464 & 601 & 75 & 1.15E-15 & 1.12E-19\\ 
	464 & 600 & 73 & 2.33E-14 & 2.97E-17\\
        \hline
    \end{tabular}
    \label{tab:BRAFboth}
} 
    
\qquad
\subfloat[Clusters found by \NMC{} and dropped by \GraphPAC{}.]{
    \begin{tabular}{|c|c|c|c|}
        \hline
        Start & End & \# Muts. & \NMC{} Pvalue              \\ \hline
	596 & 671 & 67 & 4.12E-29\\ 
	597 & 671 & 65 & 4.79E-27\\ 
	581 & 671 & 69 & 3.33E-26\\ 
	464 & 671 & 76 & 5.92E-09\\ 
	466 & 671 & 75 & 3.32E-08\\ 
	469 & 671 & 73 & 8.11E-07\\
        \hline
    \end{tabular}
	\label{tab:BRAFNMC}
}

\qquad
\caption{Table \ref{tab:BRAFboth} shows the significant clusters that were identified by both \GraphPAC{} and NMC. Table \ref{tab:BRAFNMC} shows the significant clusters that were not found to be significant under \GraphPAC{} but were found to be significant under NMC.}
\label{tab:BRAF}
\end{table}

Although it is outside the scope of this manuscript to consider every difference between Tables \ref{tab:BRAFboth} and \ref{tab:BRAFNMC}, we observe that three of the longest clusters 464-671, 466-671 and 469-671 are dropped by \GraphPAC{}. Since after alignment of the protein structural data to the mutational data (see Section \ref{data:PDB}), tertiary information was available on residues 448-603 and 610-723, these clusters cover 77.0\%, 76.3\% and 75.2\% of the all the available residues, respectively. By considering the 3D structure via \GraphPAC{}, the longest clusters are dropped and the remaining overlapping clusters focus almost exclusively on residues 464-600. 

After structure and mutation alignment, the residue substitutions in significant clusters include: G464V, G466V, G469V, G469A, N581S, G596R, L597V, LV597R, V600E, V600K, K601N and R671Q. Since R671Q does not have extensive literature and comes from a non-specified tissue sample in the COSMIC database, it will no longer be considered here. Thus, by considering the tertiary structure, we significantly narrow the window of which residues to consider for potential driver mutations and can partition the protein into three segments: I) 464-599 and II) 600 and III) 601. Segment I is primarily associated with lung and colorectal cancer as shown in \citep{Jeet_2009, naoki_missense_2002, greenman_patterns_2007, davies_mutations_2002}. Segment II represents the two most common mutations in BRAF, V600E and V600K. Overall, 95\% of BRAF mutations occur on V600, with some studies showing that V600E occurs within 73\% to 79\% of patients while V600K occurs within 12\% to 19\% of patients \citep{lovly_routine_2012,menzies_distinguishing_2012}. Mutations at this position result in the oncogene being constitutively activated with increased kinase activity and has been found in a wide range of cancers such as metastatic melanoma \citep{sosman_survival_2012}, ovarian serous carcinoma \citep{grisham_braf_2013} and hairy cell leukemia \citep{ewalt_real-time_2012}. Furthermore, recent inhibitors, such as Vemurafenib and GSK2118436 specifically target the V600E and V600E/K mutations (respectively), supporting the hypothesis that somatic clusters can provide pharmacological targets \citep{lemech_potential_2011}. Lastly, segment III is comprised of the much less common K601N mutation which has been observed in myeloma cases along with V600E. Since these patients share the more common BRAF mutations as well, they may also potentially benefit from BRAF inhibitors \citep{chapman_initial_2011}.

\section{Conclusion}

In this manuscript we provide an alternative method to utilize protein tertiary structure when identifying somatic mutation clusters. By employing a graph theoretic approach to restructuring the protein order, we identify both new clusters in proteins previously shown to have clustering as well as  proteins that were not previously shown to have clustering. We have also provided several examples where we are able to identify clusters of mutations that may benefit from pharmacological treatment. Moreover, as \GraphPAC{} uses the \NMC{} algorithm to identify clusters rather than a fixed window size, we are able to detect clusters of varying lengths. Finally, the methodology is fast and robust with the overwhelming majority of structure/protein combinations taking under 10 minutes each to analyze on a consumer desktop with an Intel i7-2600k processor running at 3.40 GHZ and 16GB of DDR3 RAM. 

The \GraphPAC{} algorithm, while presenting a viable alternative to the MDS restriction of \iPAC{} and an improvement over NMC, nevertheless contains several limitations. First, while no longer bound to the MDS requirement of \iPAC{}, there is no closed form solution to the shortest path problem and our algorithm must appeal to heuristic approximations. Further, while Figure \ref{fig:Distribution} shows that for most structures all three insertion methods are quite similar in their rearrangement of the protein, for several structures the variability between methods can still be high. As such, future research is necessary to remove the one-dimension requirement and consider the protein in its native 3D state. 

Second, to satisfy the uniformity assumption, the mutation status of all residues must be known ahead of time. With the growth of high-throughput sequencing however, this issue is temporary. Next, unequal rates of mutagenesis along with hypermutability of specific genomic regions may violate the assumption that every residue has a uniform probability of mutation. To help ensure that this assumption holds, we only consider single residue missense substitutions and have removed insertions and deletions from the analysis since they tend to be sequence dependent. Further, research has shown that CpG dinucleotides may have a mutational frequency ten times or higher compared to other dinucleotides \citep{Sved01061990}. However, in the analyses presented in Sections \ref{new}, \ref{more}, \ref{less},  only approximately 13\% of the mutations used to identify clustering occurred in CpG sites. Relatedly, colorectal carcinomas \citep{hollstein_p53_1991} contain more transition mutations while cigarette use results in more transversion mutations in lung carcinomas \citep{ye_2010}. Still, when considering KRAS, the overwhelming majority of substitutions occur on residues 12,13, and 61 for both colorectal and lung cancer, implying that while the mutational landscape may vary, it does not have a significant effect on mutation location and thus would not violate the uniformity assumption. Hence, while this analysis is influenced by a variety of factors, as are previous studies, it nevertheless appears that the primary cause of clustering is selection for a cancer phenotype. We also note that since we obtained our mutational data from COSMIC, some tissue types are more represented than others in the database. However, this scenario results in our analysis being more conservative and our findings even more significant. Assuming that mutations occur in different parts of the protein for different tissue types, when collapsing over all tissues a larger value of $n$ is obtained while the values of $i$ and $k$ (as seen in Equation \ref{eqn:continuous}) for two specific mutations are not changed. This results in a larger p-value signifying that clusters found when collapsing over tissue types would be even more significant if only a unique tissue type was analyzed.

Further, as described in Section \ref{discussion}, \GraphPAC{} finds fewer clusters for a significant percentage of the structures analyzed. Overall, the reduction in total clusters identified can result from two sources: the removal of some residues because no tertiary data was available or the cluster is no longer significant when using the traveling salesman algorithm to account for 3D structure. The first case, which is already rare, will become increasingly more so as additional studies result in more complete and detailed structural information. For the second case, if a cluster is not found to be significant under \GraphPAC{} when compared to NMC, a near or overlapping cluster is usually found (see Tables \ref{tab:BRAFboth} and \ref{tab:BRAFNMC}). For BRAF specifically, under every type of graph insertion method (cheapest, nearest and farthest), every ``probably damaging" or "possibly damaging" mutation (as classified by \emph{PolyPhen-2}) was still identified in at least one significant cluster for the structure. For a complete analysis, see ``Potential Driver Loss" in the supplementary materials. It is also worthwhile to mention that an approach that considers the protein directly in 3D space via simulation may be employed. However, such an approach would not be able to use the order statistic methodology to identify clustering and thus might not be as sensitive for small mutation counts. Additional research is required in this area.

In summary, \GraphPAC{} utilizes protein tertiary structure via a graph theoretic approach in identifying mutational clustering. We show that this method identifies new clusters that are otherwise missed and that in some cases, pharmaceutical targets for these clusters have already been found and therapies created. This helps confirm the hypothesis that mutational clustering may be indicative of driver mutations and as new protein structures become available, \GraphPAC{} can provide a rapid methodology to identify such potential mutations.

\bigskip

\section*{Author's contributions}
    GR and HZ developed the \GraphPAC{} methology. KC was responsible for obtaining the mutation data from the COSMIC database. GR and YC executed the methodology on the protein structures. GR drafted the original manuscript while KC, YC, YM and HZ were responsible for revisions. HZ finalized the manuscript. All authors have read and approved the final text.   This work was supported in part by NSF Grant DMS 1106738 (GR, HZ), NIH Grants GM59507 and CA154295(HZ).

\section*{Acknowledgements}
We thank Drs. Robert Bjornson and Nicholas Carriero for their time and help in discussing this methodology.
  
\section*{Competing Interests}
   The authors declare that they have no competing interests.

 \bibliographystyle{natbib}  
 \bibliography{article}      


\end{document}